\documentclass[a4paper,UKenglish]{lipics}

\usepackage[utf8]{inputenc}
\usepackage{microtype}
\usepackage{xspace}
\usepackage{algorithm}
\usepackage{algpseudocode}

\usepackage{placeins}

\newcommand{\LineIf}[2]{%
	\State\algorithmicif\ {#1}\ \algorithmicthen\ {#2} \algorithmicend\ \algorithmicif%
}
\newcommand{\LineWhile}[2]{%
	\State\algorithmicwhile\ {#1}\ \algorithmicdo\ {#2} \algorithmicend\ \algorithmicwhile%
}
\newcommand{\cointeger}{{\bf integer}\xspace}
\newcommand{\pluseq}{\ensuremath{\mathrel{{+}\!\!\;{=}}}}
\newcommand{\minuseq}{\ensuremath{\mathrel{{-}\!\!\;{=}}}}
\newcommand{\plusplus}{\ensuremath{{+}\!\!\;{+}}}
\newcommand{\minusminus}{\ensuremath{{-}\!\!\;{-}}}

\newcommand{\dd}{:$\!$:}
\newcommand{\Oh}{\mathcal{O}}
\newcommand{\floor}[1]{\left\lfloor\mathinner{#1} \right\rfloor}
\newcommand{\stdsort}{\texttt{std:$\!$:sort}\xspace}

     \title{
BlockQuicksort: 
How Branch Mispredictions don't affect Quicksort}
    
     \titlerunning{BlockQuicksort} 
     
     \author[1]{Stefan Edelkamp}
     \author[2]{Armin Wei\ss}
     \affil[1]{TZI, Universit{\"a}t Bremen,\\
     	Am Fallturm 1,  
     	D-28239 Bremen, Germany
     	}
     \affil[2]{Stevens Institute of Technology,\\
     	 1 Castle Point Terrace, Hoboken, NJ 07030, USA
     	}
     \authorrunning{S. Edelkamp and A. Wei\ss} 
     
     \Copyright{Stefan Edelkamp and Armin Wei\ss}
     

\begin{document}

\maketitle
\thispagestyle{empty}

\definecolor{mygreen}{rgb}{0,0.6,0}
\definecolor{mygray}{rgb}{0.5,0.5,0.5}
\definecolor{mymauve}{rgb}{0.58,0,0.82}

\begin{abstract}
Since the work of Kaligosi and Sanders (2006), it is well-known that
Quicksort~-- which is commonly considered as one of the fastest
in-place sorting algorithms~-- suffers in an essential way from branch
mispredictions.  We present a novel approach to address this problem
by partially decoupling control from data flow: in order to perform
the partitioning, we split the input in blocks of constant size (we
propose 128 data elements); then, all elements in one block are
compared with the pivot and the outcomes of the comparisons are stored in
a buffer.
In a second pass, the respective elements are rearranged. By doing so,
we avoid conditional branches based on outcomes of comparisons at
all (except for the final Insertionsort). Moreover, we prove that
for a static branch predictor the average total number of branch
mispredictions is at most $\epsilon n \log n + \Oh(n)$ for some small
$\epsilon$ depending on the block size when sorting $n$ elements.

Our experimental results are promising: when sorting random integer
data, we achieve an increase in speed (number of elements sorted per second) of more than 80\% over the
GCC implementation of C++ \stdsort. Also for many other types of data
and non-random inputs, there is still a significant speedup over
\stdsort. Only in few special cases like sorted or almost sorted inputs, 
\stdsort can beat our implementation. Moreover, even on
random input permutations, our implementation is even slightly faster
than an implementation of the highly tuned Super Scalar Sample Sort,
which uses a linear amount of additional space.
\end{abstract}

\section{Introduction}

Sorting a sequence of elements of some totally ordered universe
remains one of the most fascinating and well-studied topics in
computer science. Moreover, it is an essential part of many practical
applications. Thus, efficient sorting algorithms directly transfer to
a performance gain for many applications.  One of the most widely used
sorting algorithms is Quicksort, which has been introduced by Hoare in
1962~\cite{Hoa62} and is considered to be one of the most efficient
sorting algorithms. For sorting an array, it works as follows: first, it
chooses an arbitrary pivot element and then rearranges the
array such that all elements smaller than the pivot are moved to the
left side and all elements larger than the pivot are moved to the
right side of the array~-- this is called \emph{partitioning}. Then,
the left and right side are both sorted recursively. Although its
average\footnote{Here and in the following, the average case refers to a uniform distribution of all input permutations assuming all elements are different.} number of comparisons is not optimal~-- $1{.}38n\log n + \Oh(n)$ vs.\ $n\log n + \Oh(n)$
for Mergesort~--, its over-all instruction count is very
low. Moreover, by choosing the pivot element as median of some larger
sample, the leading term $1{.}38n\log n$ for the average number of comparisons can be made smaller~-- even
down to $n\log n$ when choosing the pivot as median of some sample of
growing size~\cite{MartinezR01}. Other advantages of Quicksort are
that it is easy to implement and that it does not need extra memory
except the recursion stack of logarithmic size (even in the worst case if properly implemented). A major drawback of Quicksort is its quadratic
worst-case running time. Nevertheless, there are efficient ways to
circumvent a really bad worst-case. The most prominent is 
Introsort (introduced by Musser~\cite{Mus97}) which is 
applied in GCC implementation of \stdsort: as soon as the
recursion depth exceeds a certain limit, the algorithm switches to
Heapsort.

Another deficiency of Quicksort is that it suffers from branch
mispredictions (or branch misses) in an essential way. On modern
processors with long pipelines (14 stages for Intel Haswell,
Broadwell, Skylake processors~-- for the older Pentium 4 processors
even more than twice as many) every branch misprediction causes a
rather long interruption of the execution since the pipeline has to be
filled anew.  In~\cite{KaligosiS06}, Kaligosi and Sanders analyzed the
number of branch mispredictions incurred by Quicksort. They examined
different simple branch prediction schemes (static prediction and
1-bit, 2-bit predictors) and showed that with all of them, Quicksort
with a random element as pivot causes on average $c n \log n + \Oh(n)$ branch mispredictions for some constant $c = 0{.}34$ (resp.\ $c = 0{.}46$, $c
= 0{.}43$). In particular, in Quicksort with random pivot element,
every fourth comparison is followed by a mispredicted branch. The
reason is that for partitioning, each element is compared with the
pivot and depending on the outcome either it is swapped with some other
element or not. Since for an optimal pivot (the median), the
probability of being smaller the pivot is $50\%$, there is no way to
predict these branches.

Kaligosi and Sanders also established that
choosing skewed pivot elements (far off the median) might even
decrease the running time because it makes branches more
predictable. This also explains why, although theoretically larger
samples for pivot selection were shown to be superior, in practice the
median-of three variant turned out to be the best.  In
\cite{BiggarNWG08}, the skewed pivot phenomenon is confirmed
experimentally. 
Moreover, in~\cite{MartinezNW15}, precise theoretical bounds on the
number of branch misses for Quicksort are given~-- establishing also theoretical superiority of skewed pivots under
the assumption that branch mispredictions are expensive.

In~\cite{BrodalM05} Brodal and Moruz proved a general lower bound on
the number of branch mispredictions given that every comparison is
followed by a conditional branch which depends on the outcome of the
comparison. In this case there are $\Omega(n\log_d n)$ branch
mispredictions for a sorting algorithm which performs $\Oh(d n\log n)$
comparisons.
As Elmasry and Katajainen remarked in~\cite{ElmasryK12}, this theorem
does not hold anymore if the results of comparisons are not used for
conditional branches. Indeed, they showed that every program can be
transformed into a program which induces only a constant number of
branch misses and whose running time is linear in the running
time of the original program. However, this general transformation introduces a 
huge constant factor overhead.
Still, in~\cite{ElmasryK12} and~\cite{ElmasryKS12} Elmasry, Katajainen
and Stenmark showed how to efficiently implement many algorithms related
to sorting with only few branch mispredictions. They call such programs 
\emph{lean}.
In particular, they present variants of Mergesort and Quicksort
suffering only very little from branch misses. 
Their Quicksort variant (called Tuned Quicksort, for details on the
implementation, see~\cite{Katajainen14}) is very fast for random
permutations~-- however, it does not behave well with duplicate
elements because it applies Lomuto's uni-directional partitioner (see
e.\,g.\ \cite{CLRS09}).

Another development in recent years is multi-pivot Quicksort (i.\,e.\
several pivots in each partitioning 
stage~\cite{dqsanalysis1,dqsanalysis3,KushagraLQM14,WildN12,dqsanalysis2}). It
started with the introduction of Yaroslavskiy's dual-pivot 
Quicksort~\cite{Yaroslavskiy09}~-- which, surprisingly, was faster than known
Quicksort variants and, thus, became the standard sorting
implementation in Oracle Java 7 and Java 8.
Concerning branch mispredictions all these multi-pivot variants behave 
essentially like ordinary 
Quicksort~\cite{MartinezNW15}; however, they have one
advantage: every data element is accessed only a few times (this is
also referred to as the number of \emph{scans}). As outlined in
\cite{dqsanalysis3}, increasing the number of pivot elements further
(up to 127 or 255), leads to Super Scalar Sample Sort, which has been
introduced by Sanders and Winkel~\cite{SandersW04}.
Super Scalar Sample Sort not only has the advantage of few scans, but
also is based on the idea of avoiding conditional branches. Indeed,
the correct bucket (the position between two pivot elements) can be
found by converting the results of comparisons to integers and
then simply performing integer arithmetic. In their experiments Sanders and
Winkel show that Super Scalar Sample Sort is approximately twice as fast
as Quicksort (\stdsort) when sorting random integer data. However,
Super Scalar Sample Sort has one major draw-back: it uses a linear
amount of extra space (for sorting $n$ data elements, it requires
space for another $n$ data elements and additionally for more than $n$
integers). In the conclusion of~\cite{KaligosiS06}, Kaligosi and Sander raised the
question:
\begin{quote}
	\emph{However, an in-place sorting algorithm that is better
          than Quicksort with skewed pivots is an open problem.}
\end{quote}
(Here, in-place means that it needs only a constant or logarithmic
amount of extra space.)  In this work, we solve the problem 
by presenting our block partition algorithm, which
allows to implement Quicksort without any branch mispredictions
incurred by conditional branches based on results of comparisons
(except for the final Insertionsort~-- also there are still
conditional branches based on the control-flow, but their amount is
relatively small). We call the resulting algorithm BlockQuicksort.
Our work is inspired by Tuned Quicksort from~\cite{ElmasryKS12}, from where
we also borrow parts of our implementation. The difference is
that by doing the partitioning block-wise, we can use Hoare's
partitioner, which is far better with duplicate elements than Lomuto's
partitioner (although Tuned Quicksort can be made working with duplicates by applying a check for duplicates similar to what we propose for BlockQuicksort as one of the further improvements in Section~\ref{sec:improvments}). 
Moreover, BlockQuicksort is also superior to Tuned Quicksort for random permutations of integers.

\subparagraph*{Our Contributions}
\begin{itemize}
	\item We present a variant of the partition procedure that
          only incurs few branch mispredictions by storing results of
          comparisons in constant size buffers.
	\item We prove an upper bound of $\epsilon n\log n + \Oh(n)$
          branch mispredictions on average, where $\epsilon < \frac{1}{16}$ for
          our proposed block size (Theorem~\ref{thm:few_branches}).
	\item We propose some improvements over the basic version.
	\item We implemented our algorithm with an \texttt{stl}-style
          interface\footnote{Code available at \url{https://github.com/weissan/BlockQuicksort}}.
	\item We conduct experiments and compare BlockQuicksort with
          \stdsort, Yaroslavskiy's dual-pivot Quicksort and
          Super Scalar Sample Sort~-- on random integer data it is faster
          than all of these and also Katajainen et al.'s
          Tuned Quicksort.
\end{itemize}

\subparagraph*{Outline}
Section~\ref{sec:prelims} introduces some general facts on branch predictors and 
mispredictions, and gives
a short account of standard improvements of Quicksort. In 
Section~\ref{sec:partition},
we give a precise description of our block partition method and
establish our main theoretical result~-- the bound on the number of
branch mispredictions. Finally, in Section~\ref{sec:experiments}, we
experimentally evaluate different block sizes, different pivot
selection strategies and compare our implementation with other state
of the art implementations of Quicksort and Super Scalar Sample Sort.

\section{Preliminaries}\label{sec:prelims}
Logarithms denoted by $\log$ are always base 2.  
The term \emph{average case} refers to a uniform distribution of all
input permutations assuming all elements are different.
In the following \stdsort always refers to its GCC implementation.

\subparagraph*{Branch Misses} Branch mispredictions can occur when the code contains conditional jumps (i.\,e.\ \emph{if} statements, loops, etc.). Whenever the execution flow reaches such a statement, the processor has to decide in advance which branch to follow and decode the subsequent instructions of that branch.
Because of the length of the pipeline of modern microprocessors, a wrong predicted branch causes a large delay since, before continuing the execution, the instructions for the other branch have to be decoded.

\subparagraph*{Branch Prediction Schemes} 
Precise branch prediction schemes of most modern processors are not
disclosed to the public. However, the simplest
schemes suffice to make BlockQuicksort induce only few
mispredictions.

 The easiest branch prediction scheme is the
\emph{static predictor}: for every conditional jump the compiler marks
the more likely branch. 
 In particular, that means that for every \emph{if} statement, we can assume that there is a
 misprediction if and only if the \emph{if} branch is not taken; for every
\emph{loop} statement, there is precisely one misprediction for every
time the execution flow reaches that loop: when the execution leaves
the loop. For more information about branch prediction schemes, we refer to \cite[Section 3.3]{HennessyP11}.

\subparagraph*{How to avoid Conditional Branches}
The usual implementation of sorting algorithms performs conditional jumps based on the outcome of comparisons of data elements. 
There are at least two methods how these conditional jumps can be
avoided~-- both are supported by the hardware of modern processors:
\begin{itemize}
	\item Conditional moves (\texttt{CMOVcc} instructions on x86
          processors)~-- or, more general, conditional execution. In
          C++ compilation to a conditional move can be (often)
          triggered by
	\begin{lstlisting}[language = C, numbers=none ]
		i = (x < y) ? j : i;
	\end{lstlisting}\vspace{-1.1cm}
	\item Cast Boolean variables to integer (\texttt{SETcc} instructions x86 processors). In C++:
\begin{lstlisting}[language = C++, numbers=none]
		int i = (x < y);
\end{lstlisting}\vspace{-0.8cm}
\end{itemize}
Also many other instruction sets support these methods (e.\,g.\ ARM
\cite{ARM11}, MIPS~\cite{MIPS95}). Still, the Intel
Architecture Optimization Reference Manual~\cite{Intel16} advises only
to use these instructions to avoid unpredictable branches (as it is the case
for sorting) since correctly predicted branches are still faster. For
more examples how to apply these methods to sorting, see
\cite{ElmasryKS12}.

\subparagraph*{Quicksort and improvements}
The central part of Quicksort is the partitioning procedure. Given
some pivot element, it returns a pointer $p$ to an element in the
array and rearranges the array such that all elements left of the $p$
are smaller or equal the pivot and all elements on the right are
greater or equal the pivot.
Quicksort first chooses some pivot element, then performs the
partitioning, and, finally, recurses on the elements smaller and larger
the pivot~-- see Algorithm~\ref{alg:QS}. We call
the procedure which organizes the calls to the partitioner the
\emph{Quicksort main loop}.
\begin{algorithm}[ht]
	\small	\caption{Quicksort}\label{alg:QS}
	\begin{algorithmic}[1]
		\Procedure{Quicksort}{$A[\ell,\dots, r]$}
		\If{$r>\ell$}
		\State  pivot $\gets$ choosePivot($A[\ell,\dots, r]$)
		\State cut $\gets$ partition($A[\ell,\dots, r]$, pivot)
		\State Quicksort($A[\ell,\dots, \mathrm{cut}-1]$)
		\State Quicksort($A[\mathrm{cut},\dots, r]$)
		\EndIf
		\EndProcedure
	\end{algorithmic}
\end{algorithm}

There are many standard improvements for Quicksort. For our optimized
Quicksort main loop (which is a modified version of Tuned Quicksort
\cite{ElmasryKS12, Katajainen14}), we implemented the following:
\begin{itemize}
	\item A very basic optimization due to Sedgewick~\cite{Sedgewick78} avoids recursion partially (e.\,g.\ \stdsort)
          or totally (here~-- this requires the introduction of
          an explicit stack).
	\item Introsort~\cite{Mus97}: there is an additional counter for
          the number of recursion levels. As soon as it exceeds some
          bound (\stdsort uses $2 \log n$~-- we use $2 \log n + 3$),
          the algorithms stops Quicksort and switches to
          Heapsort~\cite{Flo64,Wil64} (only for the respective
          sub-array). By doing so, a worst-case running time of
          $\Oh(n\log n)$ is guaranteed.
	\item Sedgewick~\cite{Sedgewick78} also proposed to switch to
          Insertionsort (see e.\,g.\ \cite[Section 5.2.1]{Knu73}) as soon
          as the array size is less than some fixed small constant (16
          for \stdsort and our implementation). There are two
          possibilities when to apply Insertionsort: either during the
          recursion, when the array size becomes too small, or at the
          very end after Quicksort has finished. We implemented the
          first possibility (in contrast to \stdsort) because for
          sorting integers, it hardly made a difference, but for
          larger data elements there was a slight speedup (in~\cite{LaMarcaL99} this was proposed as \emph{memory-tuned
            Quicksort}).
	\item After partitioning, the pivot is moved to its correct
          position and not included in the recursive calls (not applied in \stdsort).
	\item The basic version of Quicksort uses a random or fixed
          element as pivot. A slight improvement is to choose the
          pivot as median of three elements~-- typically the first, in
          the middle and the last. This is applied in \stdsort and
          many other Quicksort implementations. Sedgewick~\cite{Sedgewick78} already remarked that choosing the pivots
          from an even larger sample does not provide a significant increase
          of speed. In view of the experiments with skewed pivots~\cite{KaligosiS06}, this is no surprise. For BlockQuicksort,
          a pivot closer to the median turns out to be beneficial 
          (Figure~\ref{fig:skewed_pivots} in
          Section~\ref{sec:skewed}). Thus, it makes sense to invest more
          time to find a better pivot element. In~\cite{MartinezR01}, Martinez and Roura show that the number of
		comparisons incurred by Quicksort is minimal if the pivot element is
		selected as median of $\Theta(\sqrt{n})$ elements.  Another variant is
		to choose the pivot as median of three (resp.\ five) elements which
		themselves are medians of of three (resp.\ five) elements.  We
		implemented all these variants for our experiments~-- see
		Section~\ref{sec:pivot_experiments}.
\end{itemize}

Our main contribution is the block partitioner, 
which we describe in the next section.
\section{Block Partitioning}\label{sec:partition}
The idea of block partitioning is quite simple. Recall how Hoare's original partition procedure works (Algorithm~\ref{alg:hoare}): 
\begin{algorithm}[ht]
	\small
	\caption{Hoare's Partitioning}\label{alg:hoare}
	\begin{algorithmic}[1]
		\Procedure{Partition}{$A[\ell,\dots, r]$, pivot}
		\While{$\ell < r$}
		\LineWhile{$A[\ell] < \mathrm{pivot}$}{$\ell\plusplus$}
		\LineWhile{$A[r] > \mathrm{pivot}$}{$r\minusminus$}
		\LineIf{$\ell < r$}{ swap($A[\ell], A[r]$); $\ell\plusplus$;  $r\minusminus$}	
		\EndWhile
		\State\Return $\ell$
		\EndProcedure
	\end{algorithmic}
\end{algorithm}
Two pointers start at the leftmost and rightmost elements of the array
and move towards the middle. In every step the current element is
compared to the pivot (Line 3 and 4). If some element on the right
side is less or equal the pivot (resp.\ some element on the left side
is greater or equal), the respective pointer stops and the two
elements found this way are swapped (Line 5). Then the pointers
continue moving towards the middle.

\newcommand{\bs}{\ensuremath{B}\xspace}

The idea of BlockQuicksort (Algorithm~\ref{alg:partition}) is to separate
Lines 3 and 4 of Algorithm~\ref{alg:hoare} from Line 5: fix some block
size \bs; we introduce two buffers $\mathrm{offsets}_L[0, \dots, \bs -
  1]$ and $\mathrm{offsets}_R[0, \dots, \bs - 1]$ for storing pointers
to elements ($\mathrm{offsets}_L$ will store pointers to elements on
the left side of the array which are greater or equal than the pivot
element~-- likewise $\mathrm{offsets}_R$ for the right side). The main loop of
Algorithm~\ref{alg:partition} consists of two stages: the scanning phase
(Lines 5 to 18) and the rearrangement phase (Lines 19 to 26).

\begin{algorithm}[bth]
	\small
	\caption{Block partitioning}\label{alg:partition}
	\begin{algorithmic}[1]
		\Procedure{BlockPartition}{$A[\ell,\dots, r]$, pivot}
		\State \cointeger $\mathrm{offsets}_L[0, \dots, \bs - 1], \mathrm{offsets}_R[0, \dots, \bs - 1]$
		\State  \cointeger $\mathrm{start}_L,\mathrm{start}_R, \mathrm{num}_L,\mathrm{num}_R \gets 0$
		\While{$r - \ell + 1 > 2 \bs$}\Comment{start main loop}
		\If{$\mathrm{num}_L = 0$}\Comment{if left buffer is empty, refill it}
		\State $\mathrm{start}_L \gets 0$
		\For{$i = 0,\dots, \bs - 1$}
		\State $\mathrm{offsets}_L[\mathrm{num}_L ] \gets i$
		\State $\mathrm{num}_L \pluseq (\mathrm{pivot} \geq A[\ell + i] )$\Comment{scanning phase for left side} 
		\EndFor
		\EndIf
		\If{$\mathrm{num}_R = 0$}\Comment{if right buffer is empty, refill it}
		\State $\mathrm{start}_R \gets 0$
		\For{$i = 0,\dots, \bs - 1$}
		\State $\mathrm{offsets}_R[\mathrm{num}_R ] \gets i$
		\State $\mathrm{num}_R \pluseq (\mathrm{pivot} \leq A[r - i] )$ \Comment{scanning phase for right side}
		\EndFor	
		\EndIf
		\State \cointeger num $= \min(\mathrm{num}_L, \mathrm{num}_R)$
		\For{$j = 0,\dots, \mathrm{num} - 1$}
		\State swap($A\bigl[\ell + \mathrm{offsets}_L[\mathrm{start}_L + j]\bigr], A\bigl[r- \mathrm{offsets}_R[\mathrm{start}_R + j]\bigr]$)\Comment{rearrangement phase}
		\EndFor
		\State  $\mathrm{num}_L, \mathrm{num}_R \minuseq \mathrm{num}$; 
		$\mathrm{start}_L, \mathrm{start}_R \pluseq \mathrm{num}$ 
		\LineIf{$(\mathrm{num}_L = 0)$} {$\ell \pluseq \bs$}
		\LineIf{$(\mathrm{num}_R = 0)$} {$r \minuseq \bs$}
		\EndWhile\Comment{end main loop}
		\State compare and rearrange remaining elements
		\EndProcedure
	\end{algorithmic}
\end{algorithm}

Like for classical Hoare partition, we also start with two pointers
(or indices as in the pseudocode) to the leftmost and rightmost
element of the array.  First, the scanning phase takes place: the buffers
which are empty are refilled. In order to do so, we move the
respective pointer towards the middle and compare each element with
the pivot.  However, instead of stopping at the first element which
should be swapped, only a pointer to the element is stored in the
respective buffer (Lines 8 and 9 resp. 15 and 16~-- actually the pointer is 
always stored, but depending on the outcome of the comparison a counter holding 
the number of pointers in the buffer is increased or not) and the pointer
continues moving towards the middle.
After an entire block of $B$ elements has been scanned (either on both
sides of the array or only on one side), the rearranging phase begins:
it starts with the first positions of the two buffers and swaps the
data elements they point to (Line 21); then it continues until one of
the buffers contains no more pointers to elements which should be
swapped. Now the scanning phase is restarted and the buffer that has
run empty is filled again.

The algorithm continues this way until fewer elements
than two times the block size remain.  Now, the simplest variant is to switch
to the usual Hoare partition method for the remaining elements (in the
experiments with suffix \texttt{Hoare finish}).  But, we also can
continue with the idea of block partitioning: the algorithm scans the
remaining elements as one or two final blocks (of smaller size) and
performs a last rearrangement phase. After that, 
some elements to swap in one of the two buffers might still 
remain, while the other
buffer is empty. With one run through the buffer, all these elements
can be moved to the left resp.\ right (similar as it is done in the
Lomuto partitioning method, but without performing actual
comparisons).  We do not present the details for this final
rearranging here because on one hand it gets a little tedious and on
the other hand it does neither provide a lot of insight into the
algorithm nor is it necessary to prove our result on branch
mispredictions. 
The C++ code of this basic variant can be found in~Appendix~\ref{app:code}.

\subsection{Analysis}
If the input consists of random permutations (all data elements different), the average numbers of comparisons and swaps are the same
as for usual Quicksort with median-of-three. This is because both
Hoare's partitioner and the block partitioner preserve randomness of
the array.

The number of scanned elements (total number of elements loaded to the
registers) is increased by two times the number of swaps, because for
every swap, the data elements have to be loaded again. However, the
idea is that due to the small block size, the data elements still
remain in L1 cache when being swapped~-- so the additional scan has no
negative effect on the running time. In
Section~\ref{sec:blocksize_experiments} we see that for larger data
types and from a certain threshold on, an increasing size of the blocks
has a negative effect on the running time. Therefore, the block size
should not be chosen too large~-- we propose $B=128$ and fix this constant throughout (thus, already for inputs 
of moderate size, the buffers also do not require much more space than the 
stack for Quicksort).

\subparagraph*{Branch mispredictions}
The next theorem is our main theoretical result. For simplicity we
assume here that BlockQuicksort is implemented without the
worst-case-stopper Heapsort (i.\,e.\ there is no limit on the recursion
depth). Since there is only a low probability that a high recursion
depth is reached while the array is still large, this assumption is not
a real restriction. We analyze a static branch predictor: there is a
misprediction every time a loop is left and a misprediction every time
the \emph{if} branch of an \emph{if} statement is not taken.

\begin{theorem}\label{thm:few_branches}

Let $\mathcal{C}$ be the average number of comparisons of Quicksort with
constant size pivot sample. Then BlockQuicksort (without limit to
the recursion depth and with the same pivot selection method) with
blocksize $B$ induces at most $ \frac{6}{B}\cdot \mathcal{C} + \Oh(n)$ branch
mispredictions on average. In particular, BlockQuicksort with
median-of-three 
induces less then $ \frac{8}{B} n \log n + \Oh(n)$ 
branch mispredictions on average.
\end{theorem}
Theorem~\ref{thm:few_branches} shows that when choosing the block size
sufficiently large, the $n \log n$-term becomes very small and~-- for
real-world inputs~-- we can basically assume a linear number of branch
mispredictions. Moreover, Theorem~\ref{thm:few_branches} can be generalized to 
samples of non-constant size for pivot selection. Since the proof might
become tedious, we stick to the basic variant here.
The constant 6 in Theorem~\ref{thm:few_branches} can
be replaced by 4 when implementing Lines 19, 24, and 25 of 
Algorithm~\ref{alg:partition} with conditional moves.

\begin{proof}
First, we show that every execution of the block partitioner
Algorithm~\ref{alg:partition} on an array of length $n$ induces at most $
\frac{6}{B} n + c$ branch mispredictions for some constant $c$. In
order to do so, we only need to look at the main loop (Line 4 to
27) of Algorithm~\ref{alg:partition} because the final scanning and rearrangement
phases consider only a constant (at most $2B$) number of
elements. Inside the main loop there are three \emph{for} loops
(starting Lines 7, 14, 20), four \emph{if} statements (starting
Lines 5, 12, 24, 25) and the min calculation (whose straightforward
implementation is an \emph{if} statement~-- Line 19). We know that in
every execution of the main loop at least one of the conditions of the \emph{if}
statements in Line 5 and 12 is true because in every rearrangement phase
at least one buffer runs empty. The same holds for the two
\emph{if} statements in Line 24 and 25. Therefore, we obtain at most
two branch mispredictions for the \emph{if}s, three for the \emph{for} loops 
and one for the min in every execution of the main loop.

In every execution of the main loop, there are at least $B$
comparisons of elements with the pivot. Thus, the number of branch
misses in the main loop is at most $\frac{6}{B}$ times the number of
comparisons.
Hence, for every input permutation the total number of branch mispredictions of BlockQuicksort
is at most $ \frac{6}{B}\cdot \#\mathrm{comparisons} + (c + c')\cdot
\#\mathrm{calls\ to\ partition} + \Oh(n),$ where $c'$ it the number of
branch mispredictions of one execution of the main loop of Quicksort
(including pivot selection, which only needs a constant number of
instructions) and the $\Oh(n)$ term comes from the final
Insertionsort. The number of calls to partition is bounded by $n$
because each element can be chosen as pivot only once (since the
pivots are not contained in the arrays for the recursive calls). Thus, by taking the average over all input permutations,
the first statement follows.

The second statement follows because Quicksort with median-of-three
incurs $1{.}18n\log n + \Oh(n)$ comparisons on average~\cite{Sedgewick77}.
\end{proof}

\begin{remark}\label{rem:constant}
	The $\Oh(n)$-term in Theorem~\ref{thm:few_branches} 
	can be bounded by
	$3n$ by taking a closer look to the final rearranging phase.

We give a rough heuristic estimate: it is save to assume that the
average length of arrays on which Insertionsort is called is at least
8 (recall that we switch to Insertionsort as soon as the array size is
less than 17).
For Insertionsort there is one branch miss for each element
(when exiting the loop for finding the position) plus one for each
call of Insertionsort (when exiting the loop over all elements to
insert). Furthermore, there are at most two branch misses in the main
Quicksort loop (Lines 177 and 196 in Appendix~\ref{app:code}) for every call to
Insertionsort.  Hence, we have approximately $\frac{11}{8}n$ branch
misses due to Insertionsort.

It remains to count the constant number of branch misprediction
incurred during every call of partitioning: After exiting the main
loop of block partition, there is one more scan and rearrangement phase
with a smaller block size. This leads to at most $\leq 7$ branch
mispredictions (one extra because there is an additional case that
both buffers are empty).  The final rearranging incurs at most three
branch misses (Lines 118, 136, 140). Selecting the pivot as
median-of-three (Line 11) 
induces no branch misses since
all conditional statements are compiled to conditional moves.
Finally, there is at most one branch miss in the main Quicksort loop
for every call to partition (Line 180).
This sums up to at most $13$ branch misses per call to partition.
Because the average size of arrays treated by Insertionsort is at
least 8, the number of calls to partition is less than $n/8$.

Thus, in total the $\Oh(n)$-term in Theorem~\ref{thm:few_branches}
consists of at most $\frac{11}{8}n + \frac{13}{8}n = 3n$ branch
mispredictions.
\end{remark}

\subsection{Further Tuning of Block Partitioning}\label{sec:improvments}
We propose and implemented further tunings for our block partitioner:
\begin{enumerate}
\item Loop unrolling: since the block size is a power of two, the
  loops of the scanning phase can be unrolled four or even eight times
  without causing additional overhead.
\item Cyclic permutations instead of swaps: 
	We replace
{	\small	\begin{algorithmic}[1]
			\For{$j = 0,\dots, \mathrm{num} - 1$}
			\State swap($A\bigl[\ell + \mathrm{offsets}_L[\mathrm{start}_L + j]\bigr], A\bigl[r- \mathrm{offsets}_R[\mathrm{start}_R + j]\bigr]$)
			\EndFor
		\end{algorithmic}}
by the following code, which does not perform exactly the same data
movements, but still in the end all elements less than the pivot are
on the left and all elements greater are on the right: {
  \small \begin{algorithmic}[1] 
  	\State temp $\gets A\bigl[\ell +
      \mathrm{offsets}_L[\mathrm{start}_L]\bigr]$ 
      \State $A\bigl[\ell +
      \mathrm{offsets}_L[\mathrm{start}_L]\bigr] \gets
      A\bigl[r- \mathrm{offsets}_R[\mathrm{start}_R]\bigr]$ 
      \For{$j = 1,\dots, \mathrm{num} - 1$} 
      \State $A\bigl[r - \mathrm{offsets}_R[\mathrm{start}_R + j - 1]\bigr] \gets
    A\bigl[\ell + \mathrm{offsets}_L[\mathrm{start}_L + j]\bigr]$ 
    \State $A\bigl[\ell + \mathrm{offsets}_L[\mathrm{start}_L + j]\bigr]
    \gets A\bigl[r - \mathrm{offsets}_R[\mathrm{start}_R + j ]\bigr]$ 
    \EndFor 
    \State $A\bigl[r - \mathrm{offsets}_R[\mathrm{start}_R + \mathrm{num} - 1]\bigr] \gets$
    temp
		\end{algorithmic}}
		Note that this is also a standard improvement for partitioning~-- see e.\,g.\ \cite{AbhyankarI11}.
\end{enumerate}
In the following, we always assume these two improvements since they
are of very basic nature (plus one more small change in the final
rearrangement phase). We call the variant without
them \texttt{block\_partition\_simple}~-- its C++ code can be found in Appendix~\ref{app:code}.

The next improvement is a slight change of the algorithm: in
our experiments we noticed that for small arrays with many duplicates the
recursion depth becomes often higher than the threshold for switching
to Heapsort~-- a way to circumvent this
is an additional check for duplicates equal to the pivot if one of the following
two conditions applies:
\begin{itemize}
	\item the pivot occurs twice in the sample for pivot selection
          (in the case of median-of-three),
	\item the partitioning results very unbalanced for an array of small size.
\end{itemize} 
The check for duplicates takes place after the partitioning is completed. Only the larger half of the array is searched for elements equal to the pivot.
This check works similar to Lomuto's partitioner (indeed, we used the implementation from \cite{Katajainen14}): starting from the position of the pivot, the respective half of the array is scanned for elements equal to the pivot (this can be done by one \emph{less than} comparison since elements are already known to be greater or equal (resp.\ less or equal) the pivot)). Elements which are equal to the pivot are moved to the side of the pivot. The scan continues as long as at least every fourth element is equal to the pivot (instead every fourth one could take any other ratio~-- this guarantees that the check stops soon if there are only few duplicates).

After this check, all elements which are identified as being equal to the pivot remain in the middle of the array (between the elements larger and the elements smaller than the pivot); thus, they can be excluded from further recursive calls.
We denote this version with the suffix \texttt{duplicate check} (dc).

\section{Experiments}\label{sec:experiments}
We ran thorough experiments with implementations in C++ on different
machines with different types of data and different kinds of input
sequences. If not specified explicitly, the experiments are run on an Intel Core i5-2500K CPU (3.30GHz, 4 cores,
32KB L1 instruction and data cache, 256KB L2 cache per core and 6MB L3
shared cache) with 16GB RAM and operating system Ubuntu Linux 64bit
version 14.04.4.  We used GNU's \texttt{g++} (4.8.4); optimized with
flags \texttt{-O3 -march=native}.

For time measurements, we used \texttt{std\dd chrono\dd
  high\_resolution\_clock}, for generating random inputs, the Mersenne
Twister pseudo-random generator \texttt{std:$\!$:mt19937}. All time
measurements were repeated with the same 20 deterministically chosen
seeds~-- the displayed numbers are the average of these 20 runs.
Moreover, for each time measurement, at least 128MB of data were sorted~--
if the array size is smaller, then for this time
measurement several arrays have been sorted and the total
elapsed time measured.
Our running time plots all display the actual time divided by the
number of elements to sort on the \texttt{y}-axis.

We performed our running time experiments with three different data
types:
\begin{itemize}
	\item \texttt{int}: signed 32-bit integers.	
	\item \texttt{Vector}: 10-dimensional array of 64-bit
          floating-point numbers (\texttt{double}). The order is
          defined via the Euclidean norm~-- for every comparison the
          sums of the squares of the components are computed and then
          compared.
	\item \texttt{Record}: 21-dimensional array of 32-bit
          integers. Only the first component is compared.
\end{itemize}
The code of our implementation of BlockQuicksort as well as the other algorithms and our running time experiments is available at \url{https://github.com/weissan/BlockQuicksort}.

\subparagraph*{Different Block Sizes}\label{sec:blocksize_experiments}

\begin{figure}[ht]
  		\includegraphics{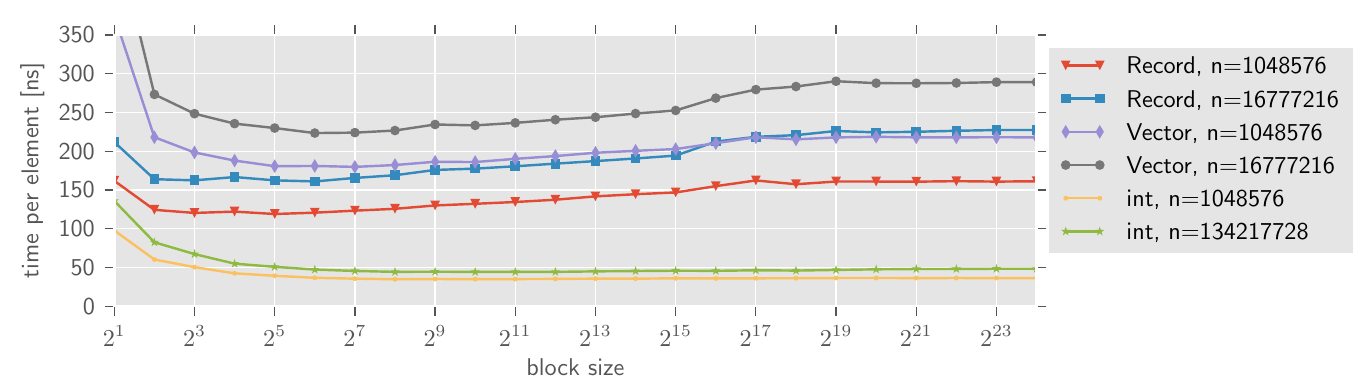}
	\caption{Different block sizes for random permutations.
}\label{fig:blocksizes}
  \end{figure}

Figure~\ref{fig:blocksizes} shows experimental results on random permutations 
for different data types and block sizes ranging from 4 up to  $2^{24}$.
We see that for integers only at the end there is a
slight negative effect when increasing the block size. Presumably this
is because up to a block size of $2^{19}$, still two blocks fit
entirely into the L3 cache of the CPU. On the other hand for
\texttt{Vector} a block size of 64 and for \texttt{Record} of 8 seem
to be optimal~-- with a considerably increasing running time for larger block 
sizes.

As a compromise we chose to fix the block size to 128 elements for all
further experiments. An alternative approach would be to choose a
fixed number of bytes for one block and adapt the block size according
to the size of the data elements.

 \subparagraph*{Skewed Pivot Experiments}\label{sec:skewed}\label{sec:pivot_experiments}
     \begin{figure}[hbt]
     	\includegraphics{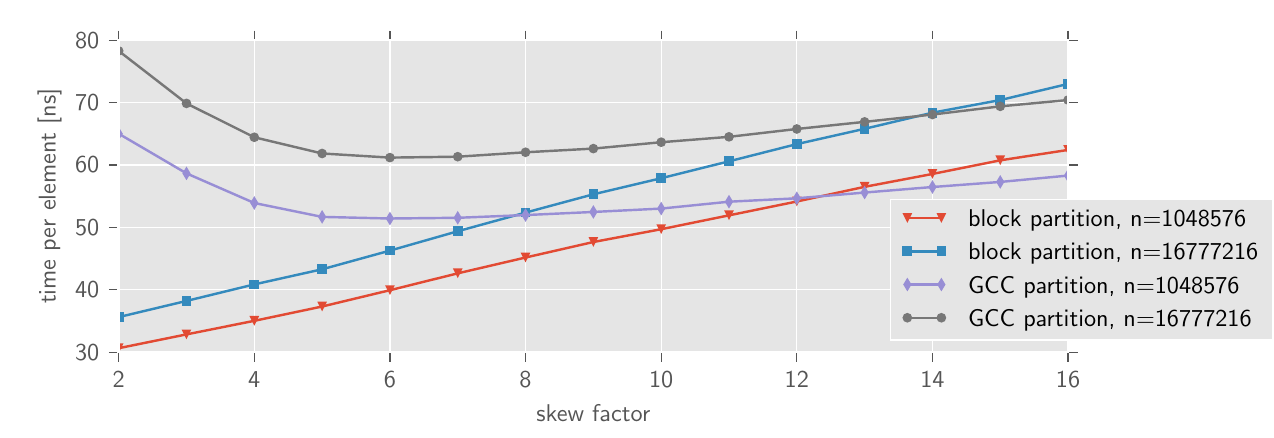}
     	\caption{Sorting random permutations of 32-bit integers with
     		skewed pivot. A skew factor $k$ means that
     		$\floor{\frac{n}{k}}$-th element is chosen as pivot of an
     		array of length $n$.}\label{fig:skewed_pivots}
     \end{figure}
We repeated the experiments from~\cite{KaligosiS06} with skewed pivot
for both the usual Hoare partitioner
(\texttt{std:$\!$:\_\_unguarded\_partition},
from the GCC implementation of \stdsort) and our block partition
method. For both partitioners we used our tuned Quicksort loop.
 The results can be seen in Figure~\ref{fig:skewed_pivots}: classic
Quicksort benefits from skewed pivot, whereas BlockQuicksort works
best with the exact median. Therefore, for BlockQuicksort it makes
sense to invest more effort to find a good pivot.

\begin{figure}[!ht]
  	\includegraphics{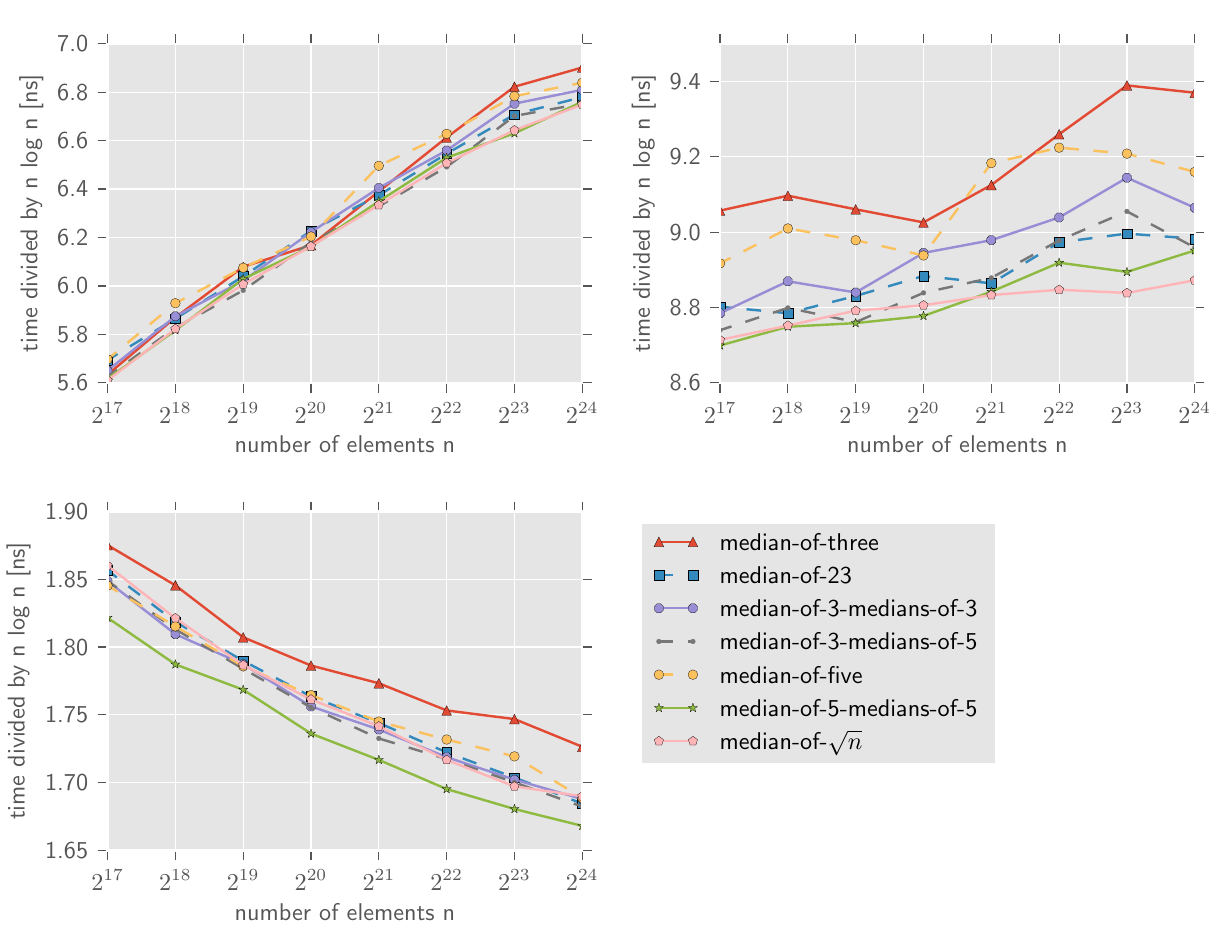}
  	\caption{Different pivot selection strategies with random permutation. 
  		\emph{Upper left}: \texttt{Record}; \emph{upper right}: 
  		\texttt{Vector}; 
  		\emph{lower left}: \texttt{int}. Be aware that the \texttt{y}-axis here 
  		displays the time divided by $n \log n$.}\label{fig:pivot_method}
  \end{figure}

  \subparagraph*{Different Pivot Selection Methods}\label{sec:pivot_selection}
  We implemented several strategies for pivot selection: 
  \begin{itemize}
  	\item median-of-three, median-of-five, median-of-twenty-three,
  	\item median-of-three-medians-of-three,
          median-of-three-medians-of-five,
          median-of-five-me\-dians-of-five: first calculate three
          (resp.\ five) times the median of three (resp.\ five)
          elements, then take the pivot as median of these three
          (resp.\ five) medians,
  	\item  median-of-$\sqrt{n}$.
  \end{itemize}
All pivot selection strategies switch to median-of-three for small
arrays. Moreover, the median-of-$\sqrt{n}$ variant switches to
median-of-five-medians-of-five for arrays of length below 20000 (for
smaller $n$ even the number of comparisons was better with
median-of-five-medians-of-five). The medians of larger samples are
computed with \texttt{std:$\!$:nth\_element}. Despite the results on
skewed pivots Figure~\ref{fig:skewed_pivots}, there was no big
difference between the different pivot selection strategies as it can be seen in Figure~\ref{fig:pivot_method}.
 As expected, median-of-three was always the slowest for larger
arrays. Median-of-five-medians-of-five was the fastest for
\texttt{int} and median-of-$\sqrt{n}$ for \texttt{Vector}. We think
that the small difference between all strategies is due to the large
overhead for the calculation of the median of a large sample~-- and
maybe because the array is rearranged in a way that is not favorable
for the next recursive calls.

\subsection{Comparison with other Sorting Algorithms}\label{sec:compare_algorithms}
We compare variants of BlockQuicksort with the GCC implementation of
\stdsort\footnote{For the source code see e.\,g.\
  \url{https://gcc.gnu.org/onlinedocs/gcc-4.7.2/libstdc++/api/a01462_source.html}~--
  be aware that in newer versions of GCC the implementation is
  slightly different: the old version uses the first, middle and last
  element as sample for pivot selection, whereas the new version uses
  the \emph{second}, middle and last element. For decreasingly sorted
  arrays the newer version works far better~-- for random permutations
  and increasingly sorted arrays, the old one is better. We used the
  old version for our experiment. The new version is included in some 
  plots Figures 9 and 10 in the appendix; this reveals a enormous difference between the two versions for particular inputs and underlines the importance of proper pivot selection. 
} 
(which is known to be one of the
most efficient Quicksort implementations~-- see e.\,g.\ \cite{BrodalFV08}), 
Yaroslavskiy's dual-pivot Quicksort~\cite{Yaroslavskiy09}
(we converted the Java code of~\cite{Yaroslavskiy09} to C++)
 and an implementation of Super Scalar Sample Sort
\cite{SandersW04} by Hübschle-Schneider\footnote{URL: \url{https://github.com/lorenzhs/ssssort/blob/b931c024cef3e6d7b7e7fd3ee3e67491d875e021/ssssort.h}~--
  retrieved April 12, 2016}. For random permutations and random values
modulo $\sqrt{n}$, we also test Tuned Quicksort 
\cite{Katajainen14} and three-pivot Quicksort implemented by Aum\"uller and  Bingmann\footnote{URL: 
  \url{http://eiche.theoinf.tu-ilmenau.de/Quicksort-experiments/}~--
  retrieved March, 2016} from~\cite{dqsanalysis3} (which is based on
\cite{KushagraLQM14})~-- for other types of inputs we omit these
algorithms because of their poor behavior with duplicate elements.

\subparagraph*{Branch mispredictions}
    \begin{figure}[htb]
   \includegraphics{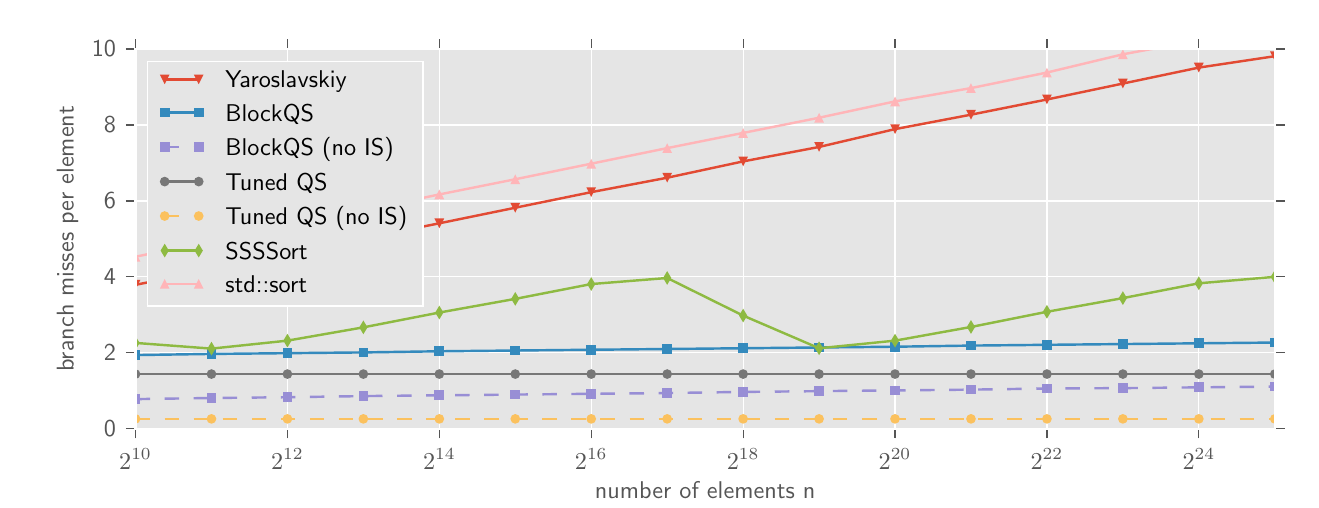}	
    	\caption{Number of branch mispredictions.}\label{fig:branch_misses}
    \end{figure}
We experimentally determined the number of branch mispredictions of
BlockQuicksort and the other algorithms with the \emph{chachegrind}
branch prediction profiler, which is part of the profiling tool
\emph{valgrind}\footnote{For more information on valgrind, see
  \url{http://valgrind.org/}. To perform the measurements we used the
  same Python script as in~\cite{ElmasryKS12,Katajainen14}, which
  first measures the number of branch mispredictions of the whole
  program including generation of test cases and then, in a second
  run, measures the number of branch mispredictions incurred by the
  generation of test cases. 
}. 
The results of these experiments on random
\texttt{int} data can be seen in Figure~\ref{fig:branch_misses}~-- the
\texttt{y}-axis shows the number of branch misprediction divided the
the array size. We only display the median-of-three variant of
BlockQuicksort since all the variants are very much alike. We also
added plots of BlockQuicksort and Tuned Quicksort skipping final
Insertionsort (i.\,e.\ the arrays remain partially unsorted).

We see that both \stdsort and Yaroslavskiy's dual-pivot
Quicksort incur $\Theta(n \log n)$ branch mispredictions. The up and
down for Super Scalar Sample Sort presumably is because of the
variation in the size of the arrays where the base case sorting
algorithm \stdsort is applied to. 
For BlockQuicksort there is an almost non-visible $n\log n$ term for the number of branch mispredictions. Indeed, we computed an approximation of $0{.}02 n \log n +	1{.}75n$ branch mispredictions.
 Thus, the actual number of branch mispredictions is still
 better then our bounds in Theorem~\ref{thm:few_branches}. There are two factors which contribute to this discrepancy: our rough estimates in the mentioned results, and that the actual branch predictor of a modern CPU might be much better than a static branch predictor. Also
note that approximately one half of the branch mispredictions are
incurred by Insertionsort~-- only the other half by the actual block
partitioning and main Quicksort loop.

Finally, Figure~\ref{fig:branch_misses} shows that Katajainen et al.'s 
Tuned Quicksort is still more efficient with respect to branch
mispredictions (only $\Oh(n)$). This is no surprise since it does not
need any checks whether buffers are empty etc. Moreover, we see that
over $80\%$ of the branch misses of Tuned Quicksort come from the
final Insertionsort. 

    \begin{figure}[tbh]
    	\includegraphics{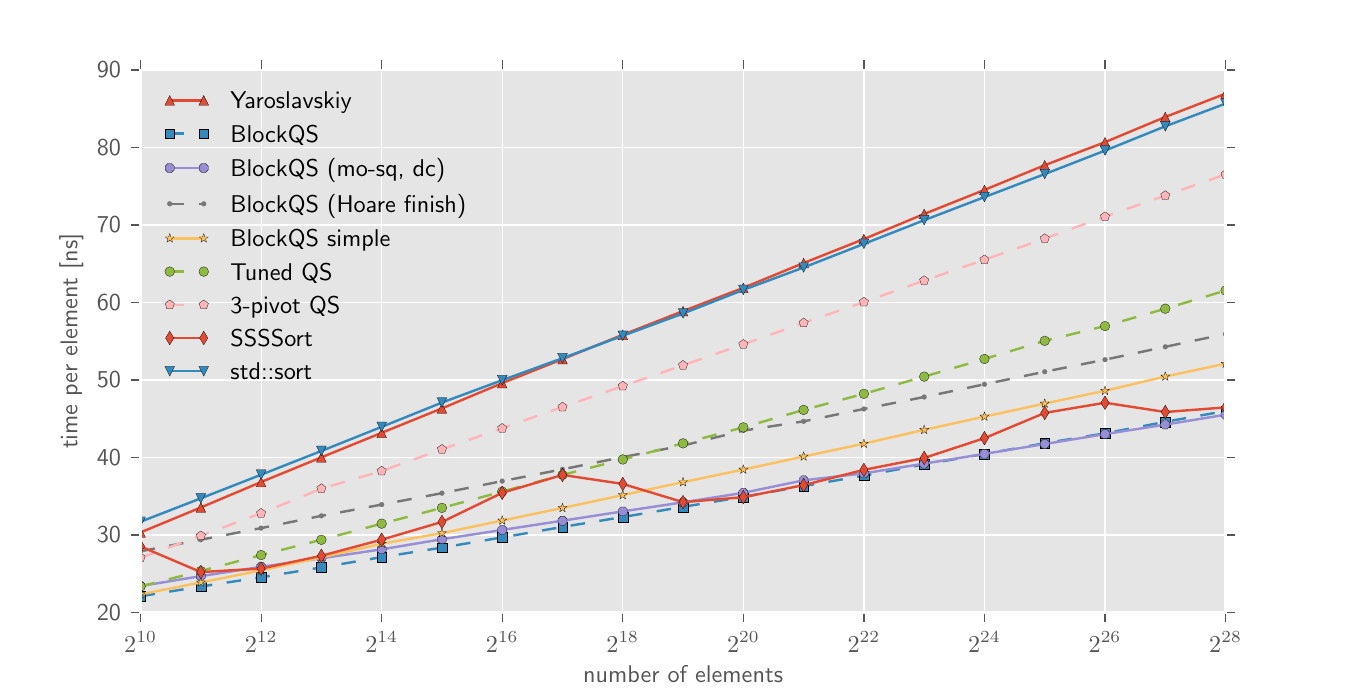}
    	\caption{Random permutation of 
    		\texttt{int}.}\label{fig:random_int}
    \end{figure}
  \subparagraph*{Running Time Experiments} 
  In Figure~\ref{fig:random_int} we present running times on random \texttt{int} 
  permutations of different BlockQuicksort variants and the other algorithms 
  including Katajainen's Tuned Quicksort and Aum\"uller and Bingmann's three-pivot Quicksort.
  The optimized BlockQuicksort variants need around 45ns per element when 
  sorting $2^{28}$ elements, whereas \stdsort needs 85ns per element~-- thus, 
  there is a speed increase of 88\% (i.\,e.\ the number of elements sorted per second is increased by 88\%)\footnote{In an earlier version of this work, we presented slightly different outcomes of our experiments. One reason it the usage of another random number generator. Otherwise, we introduced only minor changes in test environment~-- and no changes at all in the sorting algorithms themselves.}.
    \begin{figure}[!htb]
    	   	\includegraphics{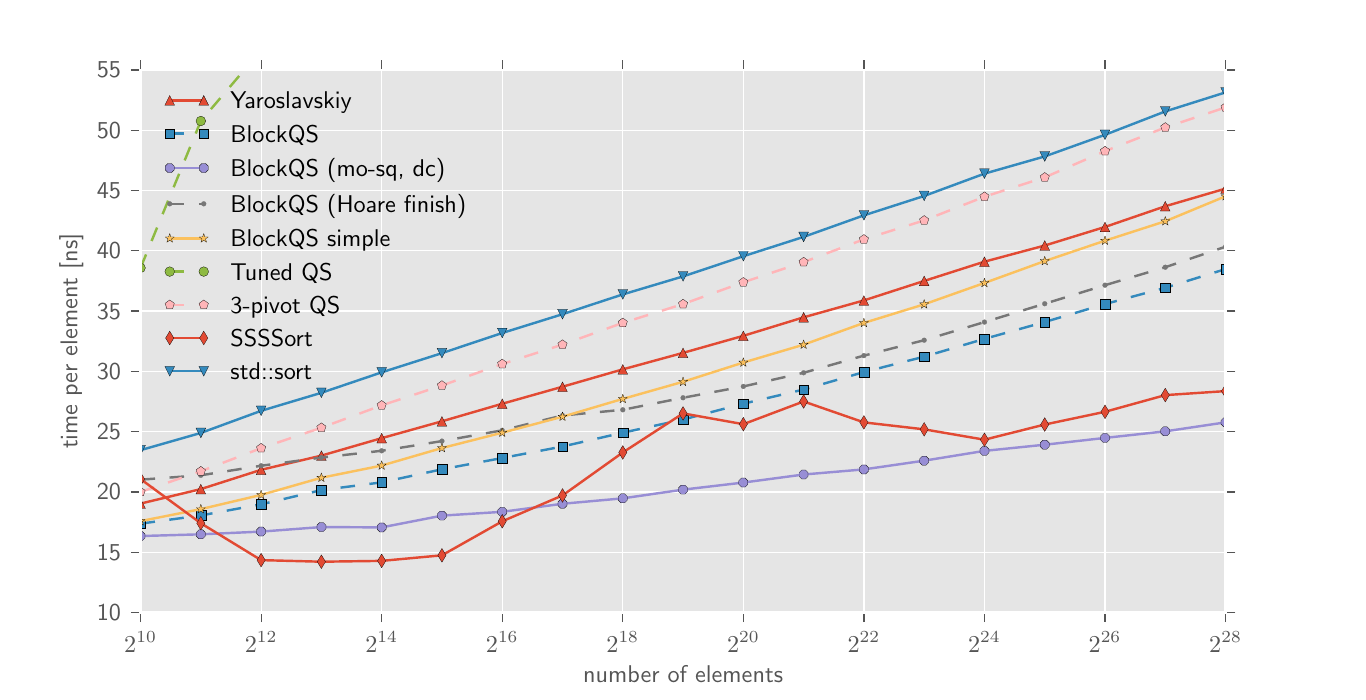}
    	\caption{Random \texttt{int} values between $0$ and 
    	$\sqrt{n}$.}\label{fig:modulo_int}
    \end{figure}
The same algorithms are displayed in Figure~\ref{fig:modulo_int} for
sorting random \texttt{int}s between $0$ and $\sqrt{n}$. Here, we
observe that Tuned Quicksort is much worse than all the other
algorithms (already for $n=2^{12}$ it moves outside the displayed range). All variants of
BlockQuicksort are faster than \stdsort~-- the
\texttt{duplicate check} (dc) version is almost twice as fast.

\begin{figure}[th]
  	\includegraphics{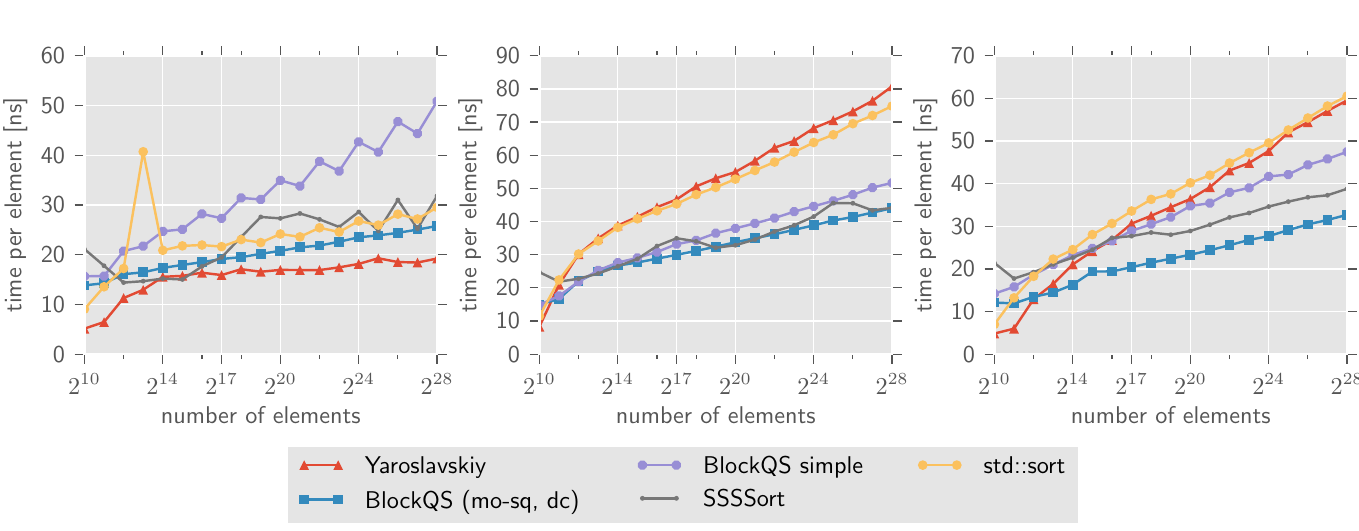}
  	\caption{Arrays $A$ of \texttt{int} with duplicates: 
  		\emph{left}: $A[i] = i \mod \floor{\sqrt{n}}$; \emph{middle}: $A[i] = i^2 + 
  		n/2\mod n$; 
  		\emph{right}: $A[i] = i^8 + n/2 \mod 
  		n$. Since $n$ is always a power of two, the value $n/2$ occurs approximately 
  		$n^{7/8}$ times in the last case.}\label{fig:powers}
  \end{figure}

Figure~\ref{fig:powers} presents experiments with data containing a lot of duplicates and having specific structures~-- thus, maybe coming closer to ``real-world'' inputs (although it is not clear what that means). Since here Tuned Quicksort and three-pivot Quicksort are much slower
than all the other algorithms, we exclude these two algorithms from the plots. The array for the left plot contains long already sorted runs. This is most likely the reason that \stdsort and Yaroslavskiy's dual-pivot Quicksort have similar running times to BlockQuicksort (for sorted sequences the conditional branches can be easily predicted what explains the fast running time). The arrays for the middle and right plot start with sorted runs and become more and more erratic; the array for the right one also contains a extremely high number of duplicates. Here the advantage of BlockQuicksort~-- avoiding conditional branches~-- can be observed again. In all three plots the check for duplicates (dc) established a considerable improvement.
  \begin{figure}[tbh]
  	\includegraphics{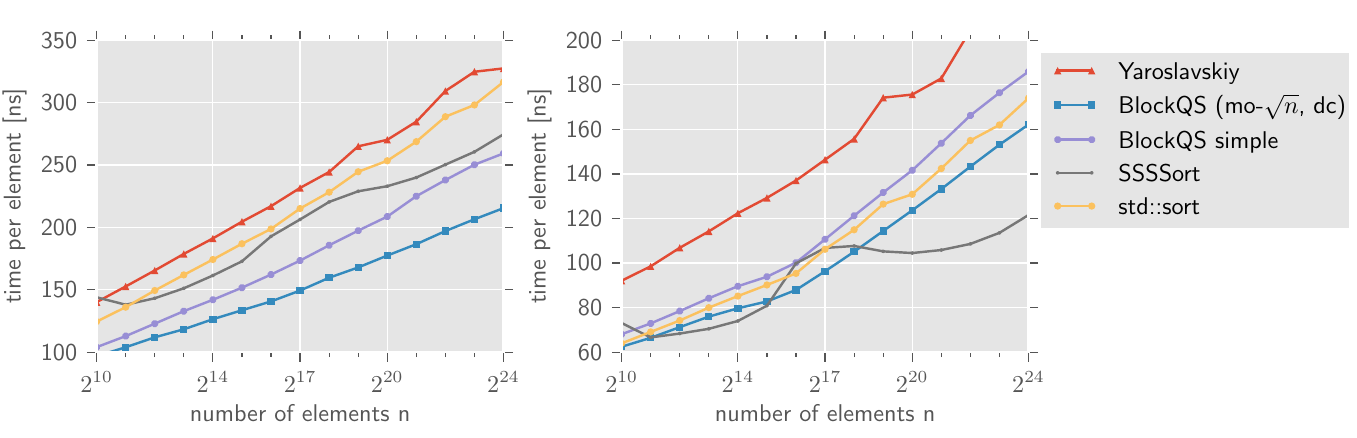}
  	\caption{Random permutations~-- \emph{left:} \texttt{Vector}; \emph{right:} 
  		\texttt{Record}.}\label{fig:RecordVector} 
  \end{figure}

In Figure~\ref{fig:RecordVector}, we show the results of selected algorithms for 
random permutations of \texttt{Vector} and \texttt{Record}. We conjecture that the good results of 
Super Scalar Sample Sort on \texttt{Record}s are because of its better 
cache behavior (since \texttt{Record} are large data elements with very cheap comparisons). More running time experiments also on other machines and compiler flags can be found in Appendix~\ref{app:more_experiments}.

  \subparagraph*{More Statistics}
  Table~\ref{tab:more_stats} shows the number of branches taken / branch mispredicted as well as the instruction count and cache misses. Although \stdsort has a much lower instruction count than the other algorithms, it induces most branch misses and (except Tuned Quicksort) most L1 cache misses (= L3 refs since no L2 cache is simulated). BlockQuicksort does not only have a low number of branch mispredictions, but also a good cache behavior~-- one reason for this is that Insertionsort is applied during the recursion and not at the very end.
  \begin{table}[!ht]
  	\begin{center}
  		\begin{small}
  			\begin{tabular}{|c||c|c||c||c|c|c|} \hline
  				algorithm & \parbox{1.2cm}{branches\\ taken} & $\vphantom{\binom{\binom{K^H}{N}}{\binom{K}{N_N}}}$\parbox{1cm}{branch\\ misses} & instructions & L1 refs & L3 refs & L3 misses \\ \hline
  				\stdsort &  37.81 & 10.23 & 174.82 & 51.96 & 1.05 & 0.41\\
  				SSSSort &  16.2 & 3.87 & 197.06 & 68.47 & 0.82 & 0.5\\
  				Yaroslavskiy &  52.92 & 9.51 & 218.42 & 59.82 & 0.79 & 0.27\\
  				BlockQS (mo-$\sqrt{n}$, dc) &  20.55 & 2.39 & 322.08 & 89.9 & 0.77 & 0.27\\
  				BlockQS (mo5-mo5) &  20.12 & 2.31 & 321.49 & 88.63 & 0.78 & 0.28\\
  				BlockQS &  20.51 & 2.25 & 337.27 & 92.45 & 0.88 & 0.3\\
  				BlockQS (no IS) &  15.38 & 1.09 & 309.85 & 84.66 & 0.88 & 0.3\\
  				Tuned QS &  29.66 & 1.44 & 461.88 & 105.43 & 1.23 & 0.39\\
  				Tuned QS (no IS) &  24.53 & 0.26 & 434.53 & 97.65 & 1.22 & 0.39\\ \hline
  			\end{tabular}
  		\end{small}
  	\end{center}
  	\caption{Instruction count, branch and cache misses when sorting random \texttt{int} permutations of size $16777216 = 2^{24}$. All displayed numbers are divided by the number of elements.}
  	\label{tab:more_stats}
  \end{table}  

\section{Conclusions and Future Research}

We have established an efficient in-place general purpose sorting algorithm,
which avoids branch predictions by converting results of comparisons
to integers. In the experiments we have seen that it is competitive on different kinds of data. Moreover, in several benchmarks it is
almost twice as fast as \stdsort.
Future research might address the following issues:
\begin{itemize}
	\item We used Insertionsort as
          recursion stopper~-- inducing a linear number of branch
          misses. Is there a more efficient recursion stopper that
          induces fewer branch mispredictions? 
	\item More efficient usage of the buffers: in our implementation the buffers on average are not even filled half. To use the space more efficiently one could address the buffers cyclically and scan until one buffer is filled.
          By doing so, also both buffers could be filled in the same loop~-- however, with the cost of introducing additional overhead.
	\item The final rearrangement of the block partitioner
          is not
          optimal: for small arrays the similar problems
          with duplicates arise as for Lomuto's
          partitioner.
	\item Pivot selection strategy: 
          though theoretically optimal, 
          median-of-$\sqrt{n}$ pivot selection 
          is not best in practice. 
          Also we want to emphasize that not only the sample size but also the selection method is 
          important (compare the different behavior of the two versions of \stdsort for sorted and reversed permutations). It might be even beneficial to use a fast pseudo-random generator (e.\,g.\ a linear congruence generator) for selecting samples for pivot selection.
	\item Parallel versions: the block structure is very well
          suited for parallelism.
	\item A three-pivot version might be
          interesting, but efficient multi-pivot variants are not
          trivial: our first attempt was much slower. 
\end{itemize}

\subparagraph*{Acknowledgments}
Thanks to Jyrki Katajainen and Max Stenmark for allowing us to use their
Python scripts for measuring branch mispredictions and cache missses
and to Lorenz Hübschle-Schneider for his implementation of 
Super Scalar Sample Sort. 
We are also indebted to Jan Philipp Wächter for
all his help with creating the plots, to Daniel Bahrdt for answering
many C++ questions, and to Christoph Greulich for his help with the
experiments.

\vfill
\pagebreak

\appendix

  \FloatBarrier
\section{More Experimental Results}\label{app:more_experiments}

\begin{figure}[!h]
        	\includegraphics{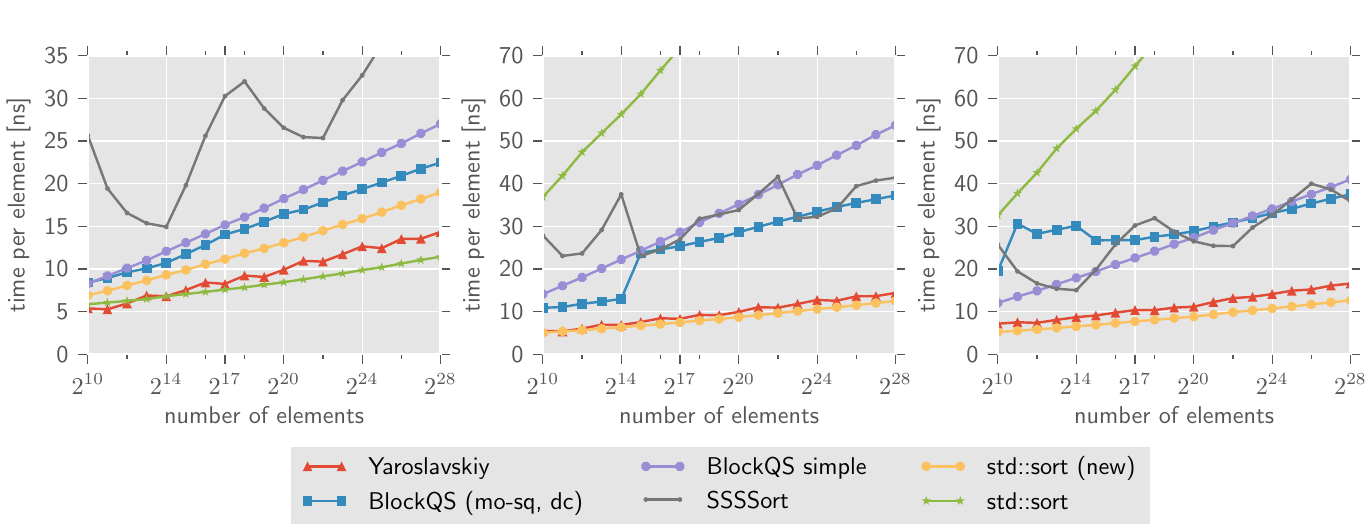}
        	\caption{Permutations of \texttt{int}: 
        		\emph{left}: sorted; \emph{middle}: reversed; \emph{right}: 
        		transposition~-- $A[i] = i + 
        		n/2 \mod n$.}\label{fig:basic_permutations}
        \end{figure}

\begin{figure}[!h]
 	\includegraphics{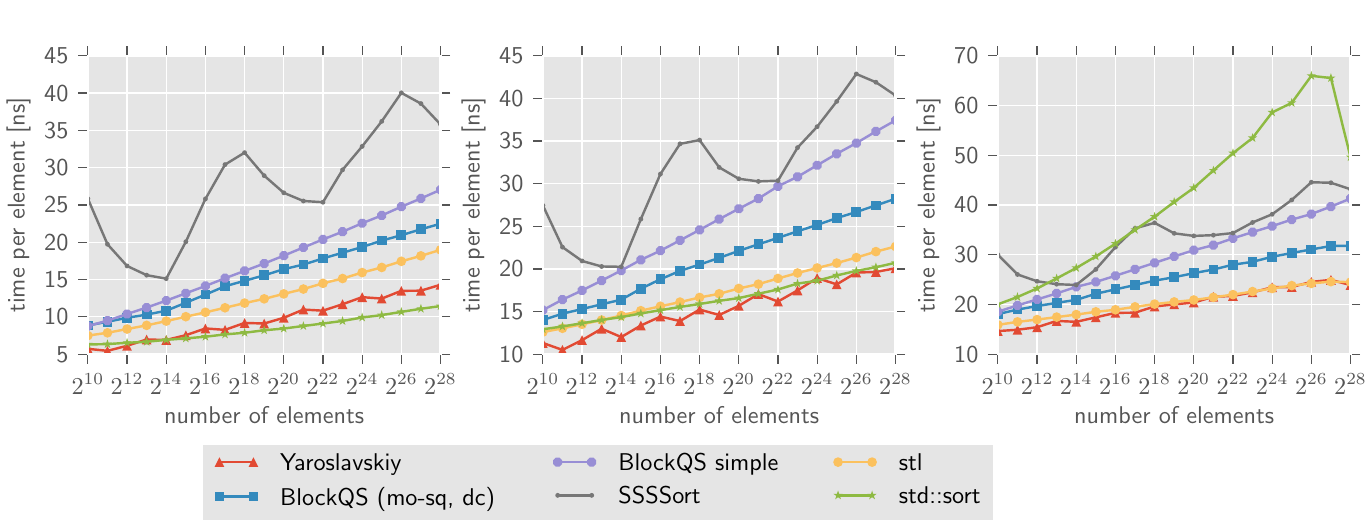}
   	\caption{Random permutations of \texttt{int} with at most $k$ 
   	inversions ($k$ random swaps of neighboring elements): 
   		\emph{left}: $k=\sqrt{n}$; \emph{middle}: $k=n$; \emph{right}: $k=n 
   		\log 
   		n$.}\label{fig:inversions}
\end{figure}

      In Figure~\ref{fig:inversions} and Figure~\ref{fig:basic_permutations} we 
      also 
      included the new GCC implementation of \stdsort (GCC version 4.8.4) 
      marked as \stdsort (new). The 
      very 
      small difference in the implementation of choosing the second instead of 
      the first element as 
      part of 
      the sample for pivot selection makes a enormous difference when sorting 
      special 
      permutations like decreasingly sorted arrays. This shows how important 
      not only 
      the size of the pivot sample but also the proper selection is. In the other benchmarks both implementations were relatively close, so we do not show both of them. 
        

\begin{figure}[ht]
	\includegraphics{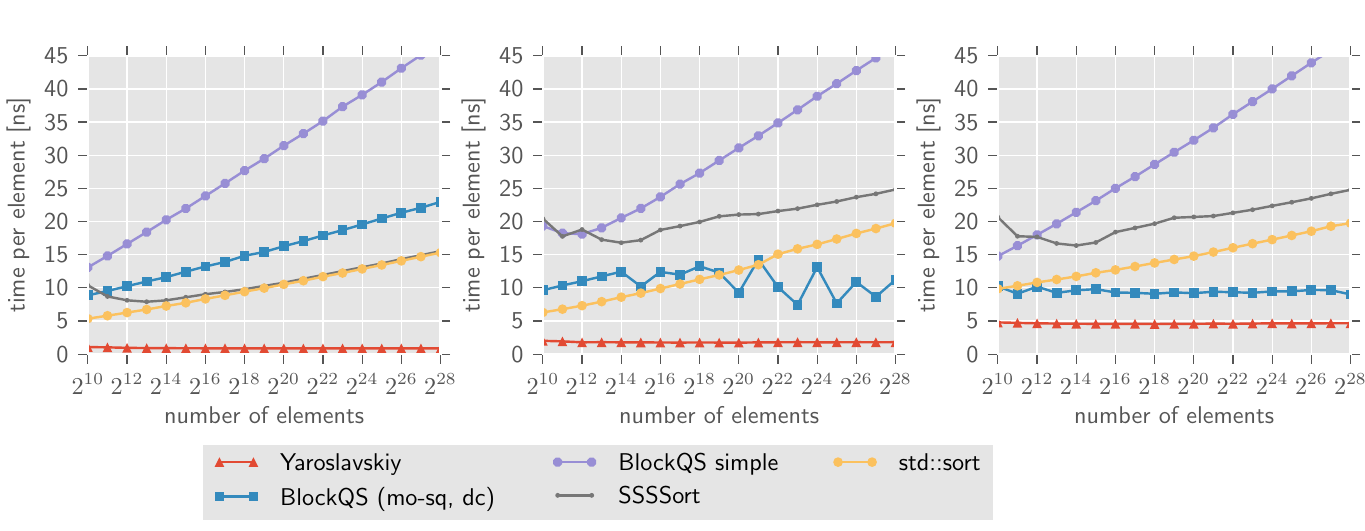}
	\caption{Arrays $A$ of \texttt{int} with many duplicates: \emph{Left:} constant; \emph{middle:} $A[i] = 0$ for $i<n/2$ and 
		$A[i] = 1$ otherwise; \emph{right:} random 0-1 values.}\label{fig:boolean}
\end{figure}

 \begin{figure}[ht]
	\includegraphics{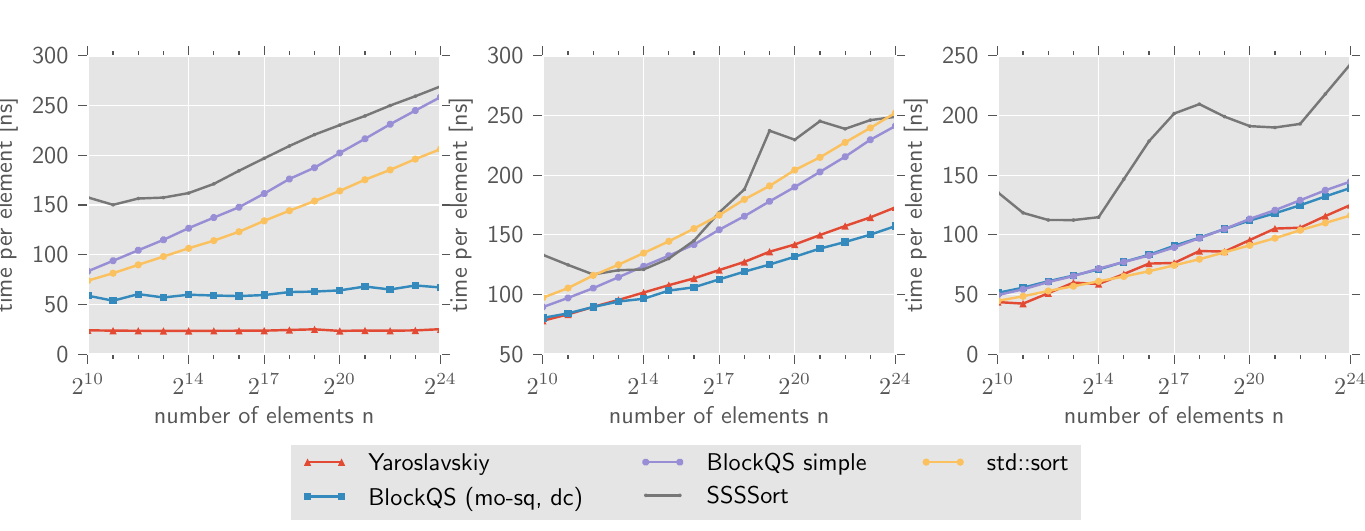}

 	\caption{ Sorting \texttt{Vector}: \emph{Left:} random 0-1 values; 
 		\emph{middle:} random values 
 		between $0$ and $\sqrt{n}$; \emph{right:} sorted.}\label{fig:Vector}
 \end{figure}

 \begin{figure}[ht]
	\includegraphics{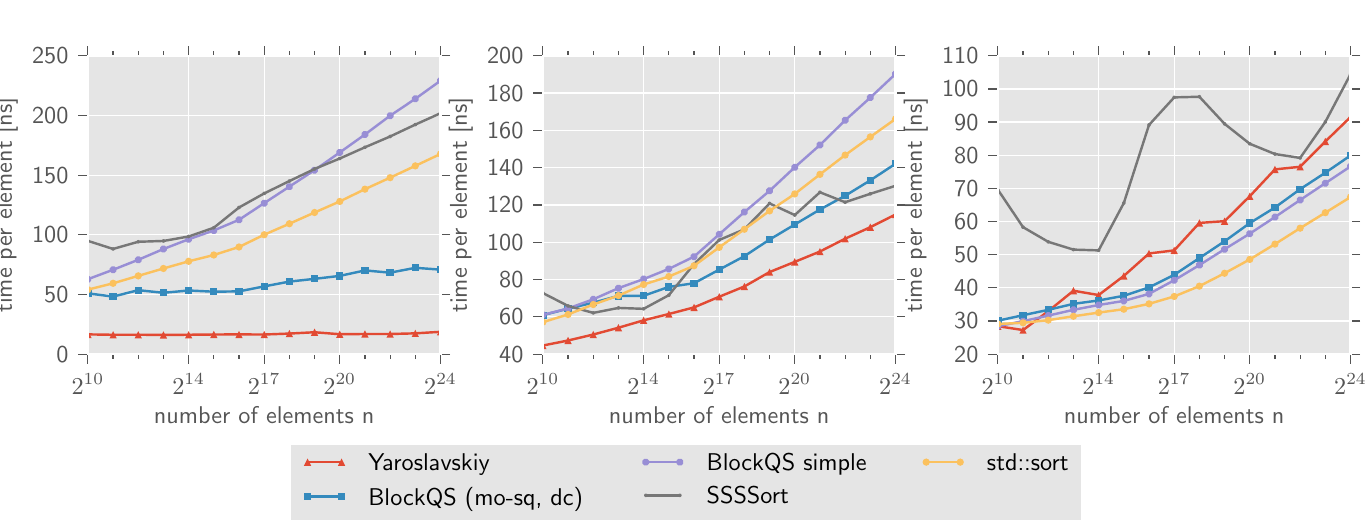}
 	\caption{ Sorting \texttt{Record}: \emph{Left:} random 0-1 values; 
 		\emph{middle:} random values 
 		between $0$ and $\sqrt{n}$; \emph{right:} sorted.}\label{fig:Record}
 \end{figure}

  \begin{figure}[bth]
	\includegraphics{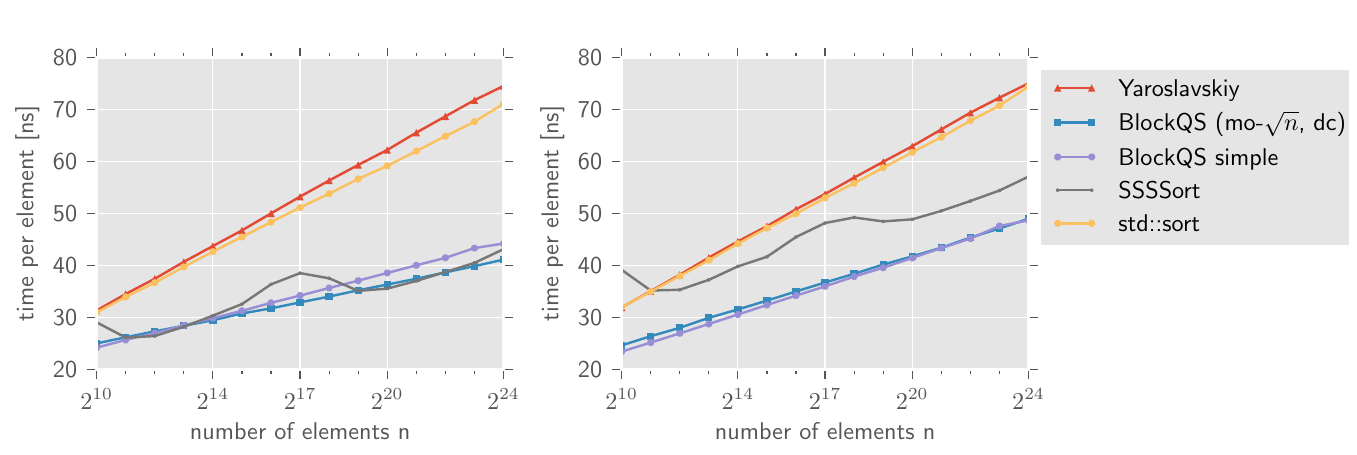}
  	\caption{Random permutations of \texttt{int} with other compiler 
  	optimizations. \\
  	\emph{Left:} \texttt{-O3 -march=native -funroll-loops}; \emph{right:} 
  		\texttt{-O1 -march=native}.}\label{fig:optimizations} 
  \end{figure}

\begin{figure}[hbt]
	\includegraphics{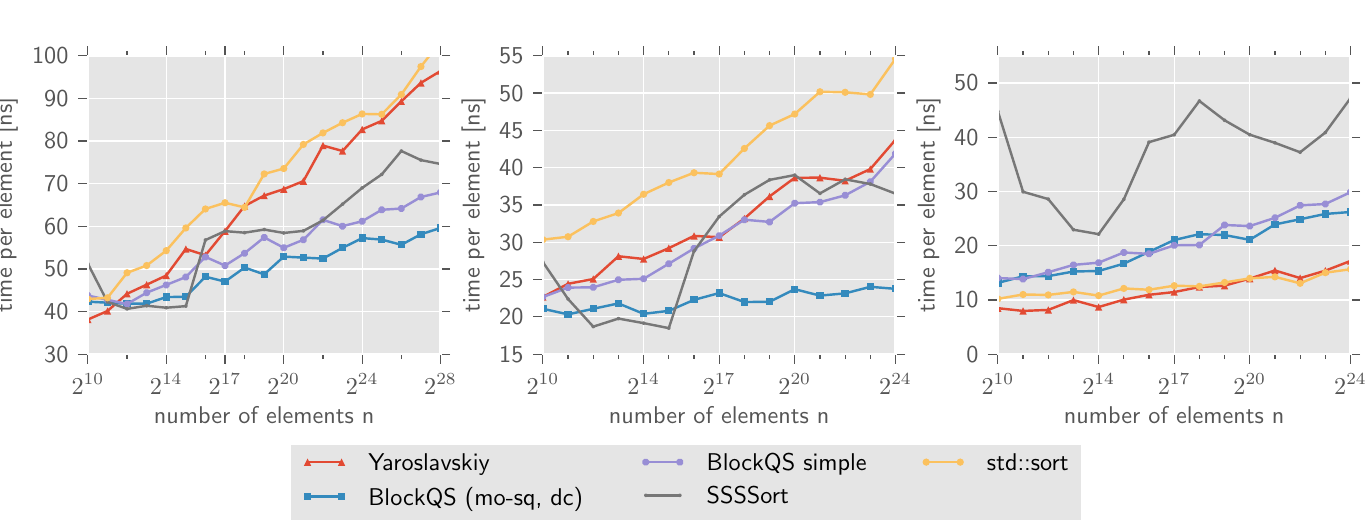}		   
      	\caption{Running time experiments with \texttt{int} on Intel Xeon  
      	E5-4627v2 CPU (3.30GHz, 8 cores, 32KB L1 instruction and data cache, 
      	256KB L2 cache per 
      	core 
      	and 16MB L3 shared cache) with 128GB RAM and operating system Windows 
      	Server 2012 R2.  We used Cygwin \texttt{g++} (4.9.3); optimized with
      		flags \texttt{-O3 -march=native}.\\ 		
      	\emph{Left}: random permutation; \emph{middle}: random 
      	values between $0$ and$\sqrt{n}$; \emph{right}: sorted.}\label{fig:xeon}
\end{figure}
     
\begin{figure}[hbt]
		\includegraphics{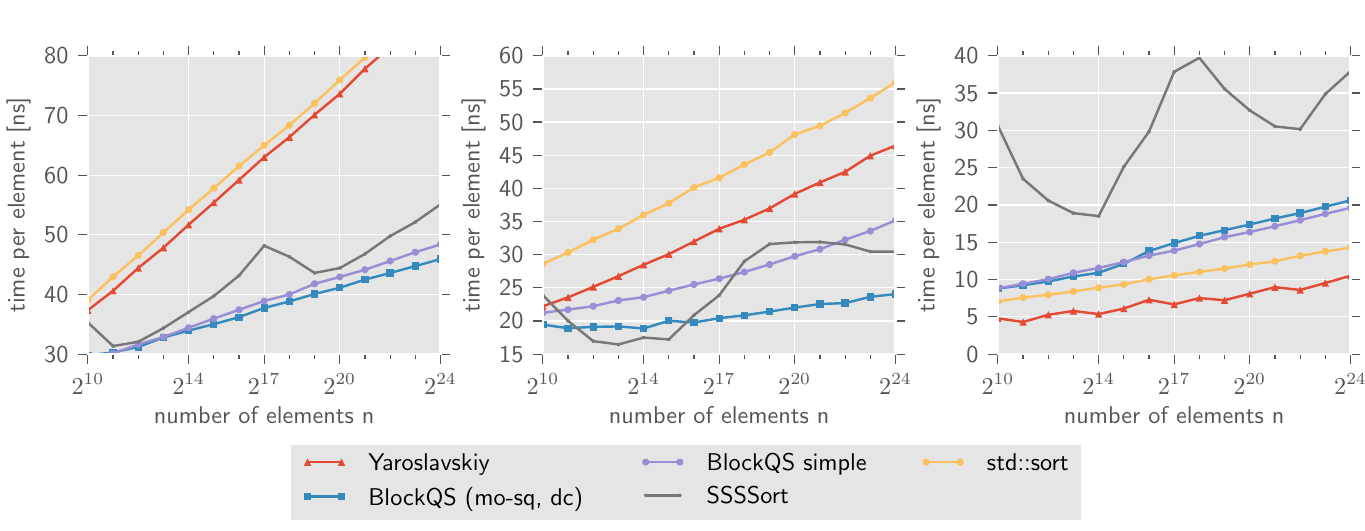}
\caption{Running time experiments with \texttt{int} on Laptop with Intel Core   
	      		i7-5500U CPU (2.40GHz, 2 cores, 32KB L1 instruction and data 
	      		cache, 256KB L2 cache per core 
	      		and 4MB L3 shared cache) with 8GB RAM and operating system 
	      		Windows 10.  We used Mingw \texttt{g++} (5.3.0); optimized 
	      		with flags \texttt{-O3 -march=native}.\\ 		
	      		\emph{Left}: random permutation; \emph{middle}: random 
	      		values between $0$ and$\sqrt{n}$; \emph{right}: 
	      		sorted.}\label{fig:laptop}
\end{figure}

    \FloatBarrier
    
\section{C++ Code}\label{app:code}
Here, we give the C++ code of the basic BlockQuicksort variant (the final 
rearranging is also in block style, but there is no loop unrolling etc. 
applied). 
\begin{lstlisting}[language = C++]
template<typename iter, typename Compare>
inline void sort_pair(iter i1, iter i2, Compare less) {
	typedef typename std::iterator_traits<iter>::value_type T;
	bool smaller = less(*i2, *i1);
	T temp = std::move(smaller ? *i1 : temp);
	*i1 = std::move(smaller ? *i2 : *i1);
	*i2 = std::move(smaller ? temp : *i2);
}
	
template<typename iter, typename Compare>	
inline iter median_of_3(iter i1, iter i2, iter i3, Compare less) {
	sort_pair(i1, i2, less);
	sort_pair(i2, i3, less);
	sort_pair(i1, i2, less);
	return i2;
}

template<typename iter, typename Compare>
inline iter hoare_block_partition_simple(iter begin, iter end, iter pivot_position, Compare less) {
	typedef typename std::iterator_traits<iter>::difference_type index;
	index indexL[BLOCKSIZE], indexR[BLOCKSIZE];
	
	iter last = end - 1;
	std::iter_swap(pivot_position, last);
	const typename std::iterator_traits<iter>::value_type & pivot = *last;
	pivot_position = last;
	last--;
	
	int num_left = 0;
	int num_right = 0;
	int start_left = 0;
	int start_right = 0;
	int num;
	//main loop
	while (last - begin + 1 > 2 * BLOCKSIZE)
	{
	//Compare and store in buffers
	if (num_left == 0) {
		start_left = 0;
		for (index j = 0; j < BLOCKSIZE; j++) {
			indexL[num_left] = j;
			num_left += (!(less(begin[j], pivot)));				
		}
	}
	if (num_right == 0) {
		start_right = 0;
		for (index j = 0; j < BLOCKSIZE; j++) {
			indexR[num_right] = j;
			num_right += !(less(pivot, *(last - j)));				
		}
	}
	//rearrange elements
	num = std::min(num_left, num_right);
	for (int j = 0; j < num; j++)
		std::iter_swap(begin + indexL[start_left + j], last - indexR[start_right + j]);
	
	num_left -= num;
	num_right -= num;
	start_left += num;
	start_right += num;
	begin += (num_left == 0) ? BLOCKSIZE : 0;
	last -= (num_right == 0) ? BLOCKSIZE : 0;
	
	}//end main loop
	
	//Compare and store in buffers final iteration
	index shiftR = 0, shiftL = 0;
	if (num_right == 0 && num_left == 0) {	//for small arrays or in the unlikely case that both buffers are empty
		shiftL = ((last - begin) + 1) / 2;
		shiftR = (last - begin) + 1 - shiftL;
		start_left = 0; start_right = 0;
		for (index j = 0; j < shiftL; j++) {
			indexL[num_left] = j;
			num_left += (!less(begin[j], pivot));
			indexR[num_right] = j;
			num_right += !less(pivot, *(last - j));
		}
		if (shiftL < shiftR)
		{
			indexR[num_right] = shiftR - 1;
			num_right += !less(pivot, *(last - shiftR + 1));
		}
	}
	else if (num_right != 0) {
		shiftL = (last - begin) - BLOCKSIZE + 1;
		shiftR = BLOCKSIZE;
		start_left = 0;
		for (index j = 0; j < shiftL; j++) {
			indexL[num_left] = j;
			num_left += (!less(begin[j], pivot));
		}
	}
	else {
		shiftL = BLOCKSIZE;
		shiftR = (last - begin) - BLOCKSIZE + 1;
		start_right = 0;
		for (index j = 0; j < shiftR; j++) {
			indexR[num_right] = j;
			num_right += !(less(pivot, *(last - j)));
		}
	}
	
	//rearrange final iteration
	num = std::min(num_left, num_right);
	for (int j = 0; j < num; j++)
		std::iter_swap(begin + indexL[start_left + j], last - indexR[start_right + j]);
	
	num_left -= num;
	num_right -= num;
	start_left += num;
	start_right += num;
	begin += (num_left == 0) ? shiftL : 0;
	last -= (num_right == 0) ? shiftR : 0;			
	//end final iteration
	
	
	//rearrange elements remaining in buffer
	if (num_left != 0)
	{
		int lowerI = start_left + num_left - 1;
		index upper = last - begin;
		//search first element to be swapped
		while (lowerI >= start_left && indexL[lowerI] == upper) {
			upper--; lowerI--;
		}
		while (lowerI >= start_left)
			std::iter_swap(begin + upper--, begin + indexL[lowerI--]);
		
		std::iter_swap(pivot_position, begin + upper + 1); // fetch the pivot 
		return begin + upper + 1;
	}
	else if (num_right != 0) {
		int lowerI = start_right + num_right - 1;
		index upper = last - begin;
		//search first element to be swapped
		while (lowerI >= start_right && indexR[lowerI] == upper) {
			upper--; lowerI--;
		}
	
		while (lowerI >= start_right)
			std::iter_swap(last - upper--, last - indexR[lowerI--]);
		
		std::iter_swap(pivot_position, last - upper);// fetch the pivot 
		return last - upper;
	}
	else { //no remaining elements
		std::iter_swap(pivot_position, begin);// fetch the pivot 
		return begin;
	}
}

template<typename iter, typename Compare>
struct Hoare_block_partition_simple {
	static inline iter partition(iter begin, iter end, Compare less) {
		//choose pivot
		iter mid = median_of_3(begin, begin + (end - begin) / 2, end, less);
		//partition
		return hoare_block_partition_simple(begin + 1, end - 1, mid, less);
	}
};

//Quicksort main loop. Implementation based on Tuned Quicksort (Elmasry, Katajainen, Stenmark)
template<template<class , class> class Partitioner, typename iter, typename Compare>
inline void qsort(iter begin, iter end, Compare less) {
	const int depth_limit = 2 * ilogb((double)(end - begin)) + 3;
	iter stack[80];
	iter* s = stack;
	int depth_stack[40];
	int depth = 0;
	int* d_s_top = depth_stack;
	*s = begin;
	*(s + 1) = end;
	s += 2;
	*d_s_top = 0;
	++d_s_top;
	do {
		if (depth < depth_limit && end - begin > IS_THRESH) {
			iter pivot = Partitioner< iter, Compare>::partition(begin, end, less);
			//Push large side to stack and continue on small side
			if (pivot - begin > end - pivot) {
				*s = begin;
				*(s + 1) = pivot;
				begin = pivot + 1;
			}
			else {
				*s = pivot + 1;
				*(s + 1) = end;
				end = pivot;
			}
			s += 2;
			depth++;
			*d_s_top = depth;
			++d_s_top;
		}
		else {
			if (end - begin > IS_THRESH)  // if recursion depth limit exceeded
				std::partial_sort(begin, end, end);
			else
				Insertionsort::insertion_sort(begin, end, less); //copy of std::__insertion_sort (GCC 4.7.2)
						
			//pop new subarray from stack
			s -= 2;
			begin = *s;
			end = *(s + 1);
			--d_s_top;
			depth = *d_s_top;
		}
	} while (s != stack);
}

//example invocation of qsort
int main(void) {
	std::vector<int> v;
	//
	//assign values to v
	//
	qsort<Hoare_block_partition_simple>(v.begin(), v.end(), std::less<int>());
}
\end{lstlisting}%
\end{document}